\documentclass[preprint2]{aastex}

\newcommand{\I}{\,{\sc i}} 
\newcommand{\II}{\,{\sc ii}} 
\newcommand{\III}{\,{\sc iii}} 
\newcommand{\IV}{\,{\sc iv}}

\newcommand{\weq}{$W_{\rm eq}$} 
\newcommand{\ebv}{$E_{B-V}$} 
\newcommand{\kms}{km~s$^{-1}$} 

\shorttitle{GR~290 in M~33\altaffilmark{1} }
\shortauthors{V. F. Polcaro et al.}

\begin{document}


\title{The Extreme LBV Star GR~290 (Romano's Star) in M~33\\
Optical Spectrophotometric Monitoring\altaffilmark{*}} 


\author{
V. F. Polcaro\altaffilmark{1}, C. Rossi\altaffilmark{2},  R. F. Viotti\altaffilmark{1}, 
S. Galleti\altaffilmark{3}, R. Gualandi\altaffilmark{3}, L. Norci\altaffilmark{4}
} 



\altaffiltext{*}{
Based on observations collected with the 4.2 m  
William Herschel Telescope of the Isaac Newton Group of Telescopes, Roque 
de los Muchachos, La Palma, Spain, the 1.52 m Cassini Telescope 
at Loiano of the Bologna Astronomical Observatory,  
and the 1.82 m Copernico Telescope at Cima Ekar, Asiago, 
of the Padova Astronomical Observatory. }
\altaffiltext{1}{Istituto di Astrofisica Spaziale e Fisica Cosmica, INAF, Roma, 
Italy \\ $email$: vitofrancesco.polcaro@iasf-roma.inaf.it  
} 
\altaffiltext{2}{Dipartimento di Fisica, Universit\`a La Sapienza, Roma, Italy}
\altaffiltext{3}{INAF, Bologna Astronomical Observatory, Loiano Observing Station, Loiano, Italy}
\altaffiltext{4}{School of Physical Sciences and NCPST, Dublin City University, 
Glasnevin, Dublin 9, Ireland }


\begin{abstract}
We study the long term, S Dor--type variability  and the present hot phase 
of the Luminous Blue Variable star GR~290 (Romano's Star) in M~33  
in order to investigate possible links between 
the LBV and WNL stages of very massive stars. 
We use intermediate resolution spectra, obtained with the William Herschel 
Telescope in December 2008, when GR 290 was at minimum ($V = \sim 18.6$), 
as well as new low resolution spectra and $B~V~R~I$ photometry obtained 
with the Loiano and Cima Ekar telescopes during 2007--2010. 
We identify more than 80 emission lines in the 3100--10000~\AA\ 
range covered by the WHT spectra, belonging to different species: 
the hydrogen Balmer and Paschen series, neutral and ionized helium, 
C~\III, N~\II--\III, S~\IV , Si~\III--\IV, and many forbidden 
lines of [N~\II], [O~\III], [S~\III], [A~\III], [Ne~\III], [Fe~\III]. 
Many lines, especially the He~\I\ triplets, show a P~Cygni profile 
with an a--e radial velocity difference from $-$300 to $-$500 \kms.  
The shape of the 4630--4713 \AA\ emission blend and of other
emission lines resembles that of WN9 stars; 
the blend deconvolution shows that the 
He \II\ 4686 \AA\ has a strong broad component with FWHM$\simeq$1700 \kms. 
During 2003--2010 the star underwent large spectral variations, 
best seen in the 4630--4686~\AA\ emission feature.
Using the late--WN spectral types of Crowther \& Smith (1997),  
GR~290 apparently varied between the WN11 and WN8--9 spectral types, 
the hotter being the star the fainter its visual magnitude.
This spectrum--visual luminosity anticorrelation of GR~290 
is reminiscent of the 
behaviour of the best studied LBVs, such as S~Dor and AG~Car. 
During the 2008 minimum we find a significant decrease in bolometric 
luminosity, which could be attributed to absorption by 
newly formed circumstellar matter. 
We suggest that, presently, the broad 4686 \AA\ line and the optical 
continuum are formed in a central WR region, while the narrow emission 
line spectrum originate in an extended, slowly expanding envelope,  
that is composed by matter ejected during 
previous high luminosity phases, and ionized by the central nucleus.  
We argue that GR~290 could have just entered in a phase  
preceeding the transition from the LBV state to late WN type. 

\end{abstract}





\section{Introduction}

In a pioneer study of luminous stars in nearby galaxies, Humphreys \& Davidson 
(1979), commenting on the evolution of the most massive stars in the Milky Way 
and the Large Magellanic Cloud, recognized that the distribution 
of the most luminous hot stars in the HR Diagram defines a locus of declining 
luminosity with decreasing temperature: the Humphreys--Davidson (HD) limit. 
Taking into account the tight upper luminosity limit observed for the yellow and 
red supergiants at log$(L/L_{\odot}) \simeq$ 5.8, these authors suggested that 
the most massive stars ($M \ge 60~ M_{\odot}$) do not evolve to cooler 
temperatures, as do stars of intermediate and low mass. 
Episodes of high mass loss, like the ones observed 
in $\eta$~Car, P~Cyg, S~Dor and the Hubble--Sandages variables in M~31 
and M~33, would be responsible for this behaviour. 
To define this group of unstable, evolved hot stars 
in the upper H--R Diagram, Conti (1984) introduced the term 
Luminous Blue Variables (LBVs). 
Later, \citet{Humphreys94} distinguished between normal LBV variability 
cycles and giant eruptions. They defined normal those cycles in which
changes of up to 1--2 magnitudes are observed in the visual band 
at more or less constant bolometric luminosity, on timescales of years to decades. 
These are the so--called {\it S~Dor variability phases}, named from the prototype 
of the class in the Large Magellanic Cloud (\citealt{Gend79,Gend01}). 
In a few cases changes of 3 magnitudes or more in the visual band 
have been recorded, 
like the ones observed for $\eta$~Car in the 19th century 
and P~Cyg in the 17th century, 
the so--called giant eruptions (\citealt{Humphreys94}). 

Since 2003, we have been carrying on an extensive monitoring of LBVs in 
M~33 (Viotti et al. 2006), mostly based on observations at the Italian Loiano and 
Cima Ekar Observatories, with the aim of investigating the physical nature and 
evolutionary status of variable stars in the upper H--R Diagram and the origin of 
their instabilities. 
Among the objects of our study, the Romano's Star (GR~290), is the most 
interesting, both for its high temperature and luminosity, 
and for the large optical variations 
(\citealt{Romano78,Sharov90,Kurtev01,Sholuk02,Polcaro03,Viotti06,Maryeva10}). 
GR~290 is an LBV placed at about 4.2 kpc to the North-East of the center of M~33, 
near the young OB association OB~89.  
Its historical light curve is characterized by ample, long--term variations 
between 16.2 and 18.2 in the $B$--band (e.g. \citealt{Romano78,Sharov90,Kurtev01, 
Sholuk02}). 
Recently, GR~290 reached a deep minimum followed by the appearance of a very hot 
spectrum, the hottest so far recorded for an LBV (Viotti et al. 2007). 
In this paper, we present new spectroscopic and photometric observations of GR~290 
collected during the present hot state. 
In \S~2 we summarize the new observations and the procedures of data analysis.  
In \S~3 we describe the intermediate resolution spectrum obtained 
with WHT in December 2008, 
and analyze the spectral variations observed during 2003--2010. 
In \S~4 we discuss our results, with some final considerations in \S~5.


\section{Observations}


This work is based on photometric and spectroscopic observations of GR~290, 
performed with several telescopes between 2003 and 2010.
The observational data taken between February 2003 and December 2006 have been 
already discussed in our previous papers 
(Polcaro et al. 2003, Viotti et al. 2006, Viotti et al. 2007). 
New low resolution spectra have been obtained 
in January 2007, January--February 2008, September 2008, February 2009 
and January 2010 with the Loiano telescope. 
All these spectra, taken with the broad wavelength range grism--4 instrumental 
setup, have a dispersion of $\sim$4 \AA\ per pixel. 
In addition, new $B$~$V$~$R$~$I$ images were obtained. 
The observations are reported in Table~1, where  
the magnitudes are the mean of two or more individual observations 
and the errors in brackets are standard deviations of the fits.    
The $BVRI$ magnitudes are derived as described in our previous papers. 
In Fig.~1 we plot all our $B$~$V$~$R$ measurements during 2003--2010. 

Intermediate resolution spectra of GR~290 have been obtained on 2008 
December 4 with the ISIS spectrograph mounted at the 4.2~m William Herschel 
telescope (WHT) of the Isaac Newton Group of Telescopes. 
The R300B and R158R gratings mounted on the blue and red arms, respectively, 
provide corresponding nominal dispersions of 0.86 \AA/pixel and 1.81 \AA/pixel.
Two exposures for each wavelength range were obtained with a 
S/N for the continuum of about 30 and 40 near 4200 \AA\ and 6500 \AA\ for 
the blue and red spectra, respectively. 
All the spectra were analyzed using the standard IRAF procedure\footnote{IRAF 
is distributed by the NOAO, which is operated by AURA under contract with NSF.}.  
Multi--Gaussian fits were used to analyze line blends and P~Cygni profiles. 
We have also made use of a blue spectrum of GR~290 obtained in September 2006 
at the WIYN 3.5~m telescope with a dispersion of 0.53 \AA/pixel (see Massey et al. 
2007), kindly provided to us by Philip Massey.




\section{The spectrum of GR~290} \label{sec:Spectrum}

\subsection{The December 2008 spectrum} \label{Hires}

The December 2008 spectrum of GR~290 is shown in Figure 2 (a—-i). 
For the line identifications, we have made use of the rich literature on the 
spectra of emission line stars, including WNL, symbiotic and B--emission stars, 
and the NIST database\footnote{ Ralchenko, Yu., Kramida, A.E., Reader, J. 
and NIST ASD Team (2008).  
NIST Atomic Spectra Database (version 3.1.5), [Online]. Available:  
http://physics.nist.gov/asd3 [2009, September 3]. National Institute  
of Standards and Technology, Gaithersburg, MD}. 
The list of the line identifications is given in Table 2.\footnote{
Table 2 is published in its entirely in the electronic edition of the 
{\it Astronomical Journal}.} 
The observed wavelength and \weq\ values listed in the table are averages 
(when available) on the two spectra for each wavelength range. 
 
Neutral helium is the atomic species most abundantly represented 
in the spectrum of GR~290. 
As can be seen from Fig.~2, all the triplet and several singlet 
transition He \I\ lines 
show a component in absorption with a velocity separation 
between the absorption and emission components  
from about $-$300 \kms\ to $-$500 \kms.  
Particularly evident is the P~Cygni profile in the 3187 and 3888 \AA\ lines 
originated from the metastable level 2$^3S$ (Fig.~2a-b). 

Near the strong 5015 \AA\ He \I\ line, about 11 \AA\ to the 
blue side, is visible an intense emission line (Fig.~2f). 
The profile of the blend is similar to that observed in September 2006 
by Massey et al. (2007) (bottom dotted spectrum in the figure),  
and by Crowther \& Smith (1997) in the WN9--11 stars of the LMC, 
where it is commonly attributed to a blend of N \II\ lines of multiplets 
19 and 22 (marked by vertical bars in the figure). 
These lines have been observed in emission in the spectrum 
of late O stars, as well as in some Be stars, 
and have been attributed to selective excitation of the upper 
3d$^3$P$^0$ and 3d$^3$F$^0$ levels (Walborn \& Howarth 2000, Walborn 2001). 
However, the line is narrower than expected if it were a blend of many 
lines in the range 4987--5007 \AA. 
In addition, the observed wavelength of the emission at about 
5003 \AA\ is also compatible with that of the [O \III] 5006.8 \AA\ nebular line, 
taking into account the stellar radial velocity. 
This, together with the presence of a weak emission near 4955 \AA\  
which could correspond to the weakest 4959 \AA\ 
component of the [O \III] doublet, seems rather to favour 
the identification of the 5003 \AA\ emission with [O \III]. 
The observed 5007/4959 intensity ratio of 4.3$\pm$2.0 is not in 
disagreement with the theoretical ratio of the [O~\III] doublet, 
although we think that both the N \II\ blend and the [O~\III] line 
do contribute by a comparable intensity to the 5003 \AA\ feature. 
Note that an emission is present around the  
expected wavelength of the auroral 7319 \AA\ transition of [O \II]. 
The presence of the [O \III] doublet emission lines during minimum 
has been recently confirmed by the higher resolution observations 
described by Maryeva \& Abolmasov (2010). 
The [O~\III] doublet has also been seen strong in emission 
in the spectrum of B416, another LBV in M~33 (Fabrika et al. 2005).  

The 4686 \AA\ Paschen--$\alpha$ line of ionized helium appears as a very strong 
symmetric emission with a peak intensity about one third of that of H$\beta$. 
This line, that is part of the broad 4600--4700 \AA\ blend, 
the so--called $f$--feature, will be discussed more in details in \S~3.2.  
Also the Paschen--$\beta$ He \II\ line at 3203 \AA\ is observed in emission, 
although not so prominently over the continuum. 
An absorption line near 4536 \AA\ is present in both the WHT blue spectra, 
as well as in the WIYN spectrum of September 2006 (Fig.~4d). 
This line can be identified with the Brackett--$\epsilon$ He \II\ 4541 \AA\ line,  
slightly blue shifted with respect to the emission line rest frame. 

The Balmer and Paschen series lines of hydrogen are seen in emission 
up to H12 and P14 at our resolution. 
The flux minimum in between the Si \IV\ 4089 \AA\ emission and  
the 4100 \AA\ H$\delta$+N~\III\ blend, that falls below 
the continuum level (Fig.~2c), has to be attributed 
to a P Cygni absorption component of H$\delta$, with a possible 
contribution of the N \III\ 4097 \AA\ line. 
The steeper blue side of H$\beta$ and H$\gamma$ also suggests the presence 
of an unresolved weak P~Cygni component (Figs.~2d, 2f). 
The near--infrared region is noisy and does not allow to resolve 
the profile of the Paschen lines (e.g., Fig.~2i).  

In addition to N \III, singly ionized nitrogen is also present with several lines, 
most of which are thought to be selectively excited (see Walborn 2001). 
C \III\ is present with a weak emission line at 5695 \AA\ and with the 
blue triplet at 4647--50--51 \AA\ with a weak peak in between 
the N \III\ 4641 \AA\ and the [Fe \III] 4658 \AA\ emission lines (see also \S~3.2).  
Si \IV\ is present with the UV (RMT~2) and blue (RMT~1) multiplets  that have the 
2$P^o$ level in common. 
Two emission lines at 4482 and 4501 \AA\ (Fig.~2d) have been identified 
with the S \IV\ recombination lines 4485.662 \AA\ and 4504.093 \AA\ 
belonging to the high excitation 2$s^2 4d ^2D-3s^2 4f ^2F^o$ transition.  
These lines have been frequently observed in the spectra of O--type stars, 
and have been first identified by Werner \& Rauch (2001). 
As discussed by these authors (see also Morrell et al. 1991), 
the presence of these lines 
might be an indication of the high intrinsic luminosity of GR~290. 

In addition to [O~\III] discussed above,
the spectrum of GR~290 displays forbidden lines of doubly ionized sulfur 
(e.g., 3797 \AA, 6312 \AA, 9532 \AA), argon (7136 \AA, 7751 \AA), 
neon (3869 \AA), and iron (3240 \AA, 5270 \AA), 
as well as the yellow and red lines of [N \II]. 
These lines are not unusual in the spectrum of luminous emission line
stars, including AG Car during its hot minimum, P~Cyg and $\eta$~Car. 
In the case of GR~290, 
they probably arise from the compact elongated (6--8 arcsec in the NS 
direction) circumstellar nebula observed by Fabrika et al. (2005) in H$\beta$, 
which has an expansion velocity of a few 10 \kms\ in the observer's direction. 
The absence of [S \II] and [Fe \II] in GR~290 is in agreement with  
its present higher ionization level than in the other LBVs. 


\subsection{The 4650 \AA\ emission blend} \label{ffeature}

Of particular interest is the strong 4640--4700 \AA\ emission feature 
that is a blend of emission lines belonging to many different species 
including of N \II, N \III, C \III, [Fe \III] and He \II\ (Fig.~3). 
We have tentatively fitted the 4620--4713 \AA\ spectral region 
with a combination of the following lines: 
N \II\ 4621.39--30.54, N \III\ 4634.14--40.64--41.85, 
C \III\ 4647.42--50.25--51.47, [Fe \III] 4658.05--67.01, He \II\ 4685.68, 
[Fe \III] 4701.53, and He \I\ 4713.15. 
For all the lines we have assumed the same FWHM of 6.2 \AA, 
uncorrected for spectral resolution.
A P~Cygni profile has been used for the He \I\ 4713 \AA\ line assuming for the 
absorption and emission components a radial velocity difference of $-$220 \kms. 
However, the fit does not fully account for the flux level around 4670 and 4690 \AA. 
We attribute this excess to the presence of broad wings of the He \II\ 
4686 \AA\ line. 
This is better seen in the lower panel of Fig.~3, where we show the spectral region 
after subtraction of the contribution of all the above emission lines except 
He \II\ 4686 \AA. 
The residual is fitted by two Gaussians with FWHM=6.2 \AA\ and 26.5 \AA\ for the 
narrow and broad components, respectively. 
The width of the latter component corresponds to
a Doppler broadening of $\sim$1000 \kms. 
The derived broad/narrow flux ratio is equal to 1.4. 
We cannot exclude a minor contribution to the 4650 \AA\ blend of other weaker 
lines, such as the C \IV\ 4658 \AA\ line, although the absence of the strong C \IV\ 
doublet at 5801--12 \AA\ would argue against its presence in the 4650 \AA\ blend. 
The final fit is shown in the top panel of Fig.~3. 
In the fit the blue tail of the 4650 \AA\ blend is attributed to N \II. 
But it could be more likely attributed to a broad component 
of the N \III\ 4634--42 \AA\ triplet. If this is the case, we estimate 
that its strength should be 30--50 $\%$ of the 4686 \AA\ broad component. 
Broad emissions could be present in other spectral regions, 
but the small contrast with the continuum and, in many cases, 
the blending of lines do not allow to perform the same analysis 
as for the He \II\ 4686 \AA\ line. 

Such a double profile, with narrow and broader emission components, has been 
observed in the Ofpe/WN9--WN10h star R99 in LMC (Crowther \& Smith 1997). 
In that case the broad component is red shifted with a FWHM of 600 \kms. 
We also recall that STIS/HST observations of $\eta$ Car, 
which have allowed to resolve 
and study the spectrum of the central source (0.1 arcsec), 
show that the central source has mainly broad (FWHM $\approx$ 850 \kms) 
permitted emission lines, while the 
strong narrow emission lines observed in the ground based spectra 
are formed in the circumstellar nebula (Hillier et al. 2001). 
It is therefore likely that a similar scenario might account for the double 
components which we find here for GR~290, 
with a broad--line, high temperature spectrum formed in an unresolved 
central source, and narrower emission lines originating in a circumstellar,
slowly expanding nebula ionized by the UV radiation of the central source. 


\subsection{Spectral variations} \label{variation}

%

The spectrum of GR~290 is generally characterized by prominent hydrogen 
and neutral helium emission lines, and by the 4630--4700 \AA\ emission blend,   
which is a feature typical of Of and WN--type stars.  
These lines have shown during the years some variability, with 
the 4630--4700 \AA\ emission blend varying the most. 
To illustrate this we have plotted in Fig.~4 the 4400--5100 \AA\ spectral range 
of GR~290 as it appeared in various observations taken during 2003-—2010. 
The spectrum of the Of/WN9 star UIT~3 in M~33 is also shown for comparison. 
In the picture, the spectra of GR~290 taken at WIYN in September 2006 and 
at WHT in December 2008 have been degraded to the Loiano spectral resolution. 
The Cima Ekar spectra of GR~290 of February and December 2004, and of UIT~3  
have a slightly lower spectral resolution than those taken 
with the Loiano telescope, so that the 4630--4660 \AA\ feature, 
the He \II\ 4686 \AA\ and the He \I\ 4713 \AA\ lines are poorly resolved. 
The Loiano January 2005 spectrum was obtained with the higher resolution grism~7. 
From this comparison we can see that, in correspondence 
with the 2006 minimum, the 4630--4700 \AA\ blend 
has become prominent and has stayed so since, with maximum strength 
in the January--February 2008 spectra. 
During 2003--2010 the strength of the hydrogen and neutral helium 
emission lines also varied, although to a lesser extent. 
These variations are summarized in Table~3 which gives the equivalent 
widths of the 4630-4700 \AA\ blend, and of the He \II\ 4686 \AA, 
He \I\ 5876 \AA\ and H$\alpha$ emission lines, with an estimated error of 10$\%$ 
for H$\alpha$ and the 4650 \AA\ blend, and 20$\%$ for the two helium lines. 
The equivalent width of the 4686 \AA\ broad$+$narrow line  
has been derived by fitting the 4630--4713 \AA\ blend in the low 
resolution spectrograms with a combination of Gaussians as described above in \S~3.2.  
From Table~3 one can see that, after February 2003, the equivalent width of 
He \II\ 4686 \AA\ has dramatically decreased till a minimum value in 2004--2005. 
At this epoch, this line was not distinguishable, at the resolution of our 
spectra, while the whole 4630-4700 \AA\ blend reached a deep minimum. 
It is possible that, during this phase, the N \II\ was the dominant contributor 
to the blend (see Polcaro et al. 2003), as it has been the case for the spectrum 
of AG Car during its Of/WN9 phase (e.g. Viotti et al. 1993, Smith et al. 1994). 
Unlike the trend shown by the 4650 \AA\ blend, 
the relative strength of H$\alpha$  
between the beginning of 2004 and 2005 displayed only a slight increase, up to 
an \weq\ of 130 \AA, and remained around this value in the following years. 
As for the He \I\ 5876 \AA\ emission line its \weq\ reached a maximum 
by the end of 2006, followed by a slight decrease in the following years. 

During mid 2006 up to January 2010 GR~290 has exhibited a spectrum  
very similar to that of late WN stars. 
Therefore in order to provide a spectral class to GR~290 during its different 
phases we have used the classification scheme for WNL stars 
proposed by Crowther \& Smith (1997). 
The strength of the N \III\ 4634--4641 \AA, He \II\ 4686 \AA\ emissions 
relative to the He \I\ lines suggests that at minimum the star belonged to 
the WN9 subtype, while before 2006, when the star was brighter, 
it could be classified WN10--11. 
The spectral type variations of GR~290 can also be analyzed considering 
the intensity of the He \II\ 4686 \AA\ and He \I\ 5876 \AA\ lines 
by plotting them in a \weq(5876) v/s \weq(4686) diagram 
used for classifying WNL stars (Crowther \& Smith, 1997). 
In this diagram, shown in Fig.~5, the areas identifying the WN and Of spectral 
types are marked and labelled. 
According to this classification scheme, GR~290 has changed spectral   
subtype from WN9 in 2003 to WN11 in 2004-2005, 
and again to WN9 since 2006. 
At the beginning of 2008 the star reached its hottest state, with 
the He \II\ 4686 \AA\ line as strong as in the LMC and galactic WN8 stars 
(Fig.~5). 
However, in our low resolution spectra of January--February 2008, there 
is no evidence for the presence in particular of a prominent N \IV\ 4067 \AA\ 
emission line which should be present, according e.g. to Crowther \& 
Smith (1997), in a spectrum of WN8 subtype, 
nor this line has been identified in the high resolution spectrum 
of January 2008 discussed by Maryeva \& Abolmasov (2010). 

Both the WIYN September 2006 and the WHT December 2008 mid resolution spectra  
of GR~290 agree qualitatively well with that of the WN9h star BE~381  
in the LMC (shown by Crowther \& Smith, 1997),  
but the emission line spectrum is definitely stronger in GR~290. 
We suggest that this effect is associated with its intrinsic luminosity,  
higher than that of the LMC and galactic WN9 stars. 
We argue that this {\it luminosity effect} can explain the 2008 position 
of GR~290 in the WN8 region in the diagram of Crowther \& Smith (1997).   
Hence, a WN9h$^+$ subtype seems more appropriate to that epoch, 
where the $plus$ sign is used in order to indicate stronger than normal 
emission lines for a WN9h spectral type.


\section{Discussion} \label{discussion} 

\subsection{The light curve} \label{lightcurve}

GR~290 has been monitored photometrically for almost 50 years. 
The historical light curve is shown in Fig.~6. 
Since 1960 GR~290 displayed luminosity minima 
in 1960--1962, 1977, 2001 and, probably, in 1986, 
all with about the same $B\sim$18.0. 
Light maxima were recorded in 1967--75 
and, probably, in 1980--85, both with $B$ around 17.2,  
and, the strongest one, in 1993--94 with $B_{max}$ = 16.2. 
The recent light curve is illustrated in Fig.~1. 
Between February 2003 and December 2004--February 2005 the star's luminosity 
gradually increased by about half a magnitude up to a maximum near 2005.0  
with $B \simeq 17.1$.  
This was about one mag fainter than in the 1993 maximum,
but comparable to the two previous maxima.    
In November 2006, a marked luminosity decrease 
was recorded in all bands, with a magnitude jump of about +1.3 in all colors.  
Since then the star has remained at minimum with small photometric 
variations. 
Mid infrared observations of GR~290 with the SPITZER satellite,
reported by McQuinn et al. (2007), showed a trend similar to that 
of the optical bands, with a slight 
increase in the 3.6 $\mu$m and 4.5 $\mu$m bands between January 2004 and 
January 2005, followed in August 2005 by a large flux decrease of +0.6 mag 
in both bands.  
A flux decrease in the blue by the end of 2005 was also recorded by 
Maryeva \& Abolmasov (2010). 
These observations suggest that the star's fading has started in 2005.
Since, according to Massey et al. (2007) 
and to Maryeva \& Abolmasov (2010), in August--September 2006 the spectrum 
of GR~290 already displayed a prominent 4650 \AA\ emission blend 
similar to the present one, 
we argue that at that date the decrease to minimum had already completed. Hence, 
the star must have faded at a rate of $\ge+0.10$ mag per month,  
apparently faster than the previous fading phases (e.g. Sholukhova et al. 2002). 

The large photometric variation of 2006 was accompanied 
by a profound spectroscopic evolution, 
which is illustrated in Fig.~4 and quantified in Table~3. 
There is a clear opposite trend between the visual brightness and the equivalent 
widths of the 4630--4700 \AA\ blend and of the He \II\ 4686 \AA\ line. 
This is best illustrated in Fig.~7 where the equivalent width 
of these high temperature features is plotted against the visual magnitude. 
The {\it hot phase} corresponds to the \weq\ value of $\sim$40 \AA\ 
for the 4630--4700 \AA\ blend, when the visual magnitude was about 18.5 
with small photometric variations. 
The single He \II\ 4686 \AA\ line contributed then $\approx$40$\%$ to the blend.
This blend became about four times weaker during the high luminosity 
($V \sim$17.2--17.7) phase. This plot seems to suggest a physical 
correlation between \weq\ 
and $V$, although the lack of observations at intermediate visual magnitudes 
prevents a quantitative evaluation. 
As for the H$\alpha$ and He \I\ 5876 \AA\ emission lines, 
both appear to have slightly strengthened during the low luminosity
phase, but this increase 
seems to be less correlated with the visual magnitude of the star. 

Two spectra of GR~290 taken in September 1998 and July 1999 at the 6~m 
BTA telescope during the descending luminosity phase, are described 
by Fabrika et al. (2005). 
These authors identified prominent Balmer and He \I\ emission lines 
without significant emission in the 4650  \AA\ blend. 
At that epoch, the star had a luminosity 
of about $B \sim 17$--$17.6$ (see Sholukhova et al. 2002). 
According to Szeifert (1996), a spectrum taken in October 1992 
near the strong 1992 maximum ($B \sim 16.2$), showed, 
in addition to prominent H${\alpha}$, 
faint He \I\ and a few metal lines (see also top of Fig.~1 in Fabrika 2000). 
This led Szeifert to suggest a late--B spectral type for GR~290 in 1992. 
These earlier observations confirm and extend the above discussed 
visual luminosity--spectrum counter trend. 

It is known that in the LBVs, when they undergo ample, long--term photometric 
S~Dor type variations, the fading in the visual is accompanied 
by an increase of the excitation temperature of the emission line spectrum 
and, in some cases, by the blueing of the color index. 
The {\it spectrum-visual luminosity anti-correlation} observed in GR~290 
is reminiscent of the behaviour of the best studied LBVs, such as S~Dor and 
R127 in LMC, and the galactic object AG~Car.
In this regard, GR~290 is peculiar for the excitation 
temperature reached during its minimum phase, 
one of the highest ever observed in an LBV, 
if we exclude the explosive, LBV--like behaviour 
of one stellar component of the massive close binary system HD~5980.  
This star seems to have displayed a WNE spectrum during the long lasting phase, 
prior to the 1993--94 outbursts, when its spectrum became WN11--B1.5  
(Koenigsberger 2004, Koenigsberger et al. 2010).
No significant blueing of the color index of GR~290 was observed, 
but this could be attributed to the fact that during our period of observations, 
the star always displayed a peculiar hot spectrum 
with an energy distribution likely far from that of normal early 
type stars. 
Although, admittedly, our color index measurements  
can be inaccurate for such a faint object. 
%


\subsection{How luminous is GR~290?} \label{luminosity}
In order to put GR 290 in the context of the other known LBVs 
it is necessary to estimate the star's luminosity in its various phases.
The black--body fit of the optical-–near infrared energy distribution of GR~290 
in December 2004, when the star had $V$=17.2 and a WN11--type spectrum, 
provides a black--body temperature between 20\,000 and 30\,000~K (Viotti et al. 2006). 
The large range is due to the uncertainty on the adopted color excess, 
respectively \ebv = 0.16 (the average of the nearby associations OB~88 and OB~89) 
and \ebv = 0.22 (assuming for the unreddened $U-B$ and $B-V$ the color indices 
as for late--O stars). 
Crowther \& Smith (1997), from the analysis of the late WN stars in the LMC, 
derived for the WN11 spectral type an effective
temperature of 25\,000--27\,000~K and a bolometric correction around $-$2.8. 
Groh et al. (2009) derived $T_{eff}$=22\,800~K and $BC$=$-$2.5  
for the galactic LBV AG~Car during its 1985--1990 visual minimum, 
when the star exhibited a WN11--type spectrum. 
If we adopt for the December 2004 spectrum of GR~290 the bolometric
correction of AG Car during its WN11 phase and  
the same color excess \ebv = 0.16 of the nearby OB associations, 
we derive $M_{bol}$= $-$10.6, or $L_{bol}$ = 1.4$\times$10$^6$~$L_{\odot}$, 
with an assumed distance modulus for M~33 of 24.8 (from Kim et al. 2002). 
Then, if we assume for GR~290 $T_{eff}$=22\,800~K, 
we derive an effective radius $R_{eff} \simeq$76~$R_{\odot}$. 

At the February 2008 deep minimum the star was about 1.5 mag fainter 
in $V$ than in December 2004, 
while its spectrum was intermediate between WN9 and WN8. 
Assuming for this phase a bolometric correction of 
$-$3.0/$-$3.3 (e.g., Nugis \& Lamers 2000), and the same \ebv\ as above, 
we obtain $M_{bol} \simeq -$9.8, or  
$L_{bol} \simeq$ 0.66$\times$10$^6$~$L_{\odot}$, 
therefore a bolometric luminosity which is about half than in December 2004. 
Taking an effective temperature suitable for WN9 stars, $\sim$28\,000~K, 
we obtain an effective radius of $\sim$35~$R_{\odot}$. 

At the highest maximum of 1993 GR~290 was 0.9 mag brighter in $B$ 
than in December 2004. 
According to Szeifert (1996) the line excitation was much 
lower, probably similar to that of AG~Car during rise to a visual maximum. 
If we assume a visual bolometric correction of $-$1.2, similar to that  
derived by Groh et al. (2009) for AG~Car near maximum, 
and a reddening corrected $B-V$ color index close to zero, 
a bolometric magnitude around $-$10.5 
is obtained, close that of December 2004. 
The corresponding effective radius is as large 
as about 190~$R_{\odot}$ for a $T_{eff}$ value of 14\,000~K. 
Of course, one should keep in mind that the value of $M_{bol}$  
obtained for 1993 is quite uncertain, 
mostly due to the large uncertainty on the adopted $BC$, 
and we cannot exclude a 1993 luminosity much different from that of December 2004. 

If we follow what is generally known for LBV behaviour, in GR~290 the counter 
trend of the visual luminosity and of the emission line excitation 
would suggest a variation with more or less constant bolometric luminosity, 
similarly to what has been inferred for the S~Dor variation of AG~Car (e.g. 
Viotti et al. 1984, Lamers et al. 1989, Leitherer et al. 1994). 
However, if we consider the energy distribution of GR~290 
at the minimum visual luminosity of 2008, it is not so easy to justify how the 
bolometric luminosity might have remained constant since December 2004. 
Of course, our luminosity estimates critically depend on the assumed values of the
bolometric corrections.  
According to the proposed model atmospheres as also discussed above, 
the range of the uncertainty on the $BC$ difference between the two epochs 
is likely to be around $\pm$0.3 magnitudes. 
However, even taking this uncertainty into account, 
the luminosity difference of about a factor two 
cannot be attributed to a change of the energy distribution 
at constant luminosity alone, since it would imply an unlikely, too negative 
bolometric correction for the February 2008 spectrum. 
We suggest that this bolometric luminosity decrease could be only apparent, and  
due to a $\sim$0.8 mag extinction of the visual light, 
additional to the interstellar one. 
In this hypothesis, the light from GR~290, in February 2008, is partly absorbed  
by a circumstellar opaque envelope formed following the 2004--2005 light maximum, 
fed by matter ejected by the star. 
It would be interesting to find out whether this apparent luminosity 
decrease is associated with an increase of the mid--infrared flux. 

We finally remark that, although it is difficult to interpret in the light 
of current models a change in $L_{bol}$ during S~Dor variations for LBV stars, 
other authors have made similar suggestions e.g. for S~Dor (van Genderen et al. 1997),  
AG~Car (Groh et al. 2009) and AFGL2298 (Clark et al. 2009). 
According to Groh et al. (2009) this result for AG ~Car would imply the presence 
of physical mechanisms of conversion of the stellar radiative power into 
mechanical power to expand the outer layers. 
Previously, a similar mechanism was proposed 
by Andriesse, Donn \& Viotti (1978) for $\eta$ Car 
to relate a $+$1 mag decrease in bolometric luminosity since 1840  
to the excess of mechanical power needed to drive its massive stellar wind.  
In the case of GR~290, however, so far there are no independent evidences, 
such as a very high mass loss rate, 
that such a mechanism is at work after its 2005 maximum. 
Whether the intrinsic luminosity of GR~290 has changed during its 
S~Dor variations is a point which will require further analysis. 


\subsection{The emitting envelope} \label{envelope}

We have seen that, in addition to the narrow He \I\ emission lines 
with a P~Cygni profile, broad and narrow components are present 
in the He \II\ 4686 \AA\ line with comparable strength.  
The broad component has been identified in the mid resolution spectrum 
of GR~290 obtained in December 2008, during the low luminosity phase 
of the star. 
Similar broad wings are not seen in the prominent hydrogen lines. 
The He \II\ broad component indicates the presence of a high velocity 
($\sim$1000~\kms) region, hotter than the region 
producing the narrow emission line spectrum. 
As for the latter, 
the present day narrow line spectrum mimics fairly well the spectrum of a WN9 star, 
with a wide ionization range and a low expansion velocity (of a few 100~\kms). 

The presence of two - high and low velocity - spectral components 
might be explained by line formation in a bimodal stellar wind, 
with the broad He \II\ component generated in a fast, hot polar flow, 
and the narrow lines in an equatorial, denser and cooler one. 
Such bimodal winds have been invoked to explain similar observations 
in other LBVs. 
A wind asymmetry could also be suggested by the asymmetric shape 
of the circumstellar nebula observed by Fabrika et al. (2005). 
Anyhow, the actual spatial structure of the nebula requires a better 
assessment with higher resolution obsservations, as its shape might provide 
information about the past history of the star. 
%
We instead suggest that the hot region could be identified with the central 
stellar nucleus of GR~290 with a dense high velocity wind, similar to that of 
Wolf--Rayet stars, surrounded by a cooler shell which is expanding 
with a low velocity. 
This shell, fed by the stellar wind, would have higher optical thickness 
during the high luminosity phases. 
The continuum therefore forms in the envelope  
at different apparent radii in different luminosity phases. 
During the minimum phases, the envelope opacity would become lower and 
the spectrum of the underlying nucleus emerges. At this time 
the measured radius would correspond to  the WR--type envelope of the central nucleus. 
Other broad emission lines in addition to that of  He \II\ 4686 \AA\ that should be formed 
in the central nucleus, could be masked by the rich narrow emission line spectrum. 
This point will be analyzed with new higher quality spectra. 


\subsection{Evolutionary considerations} \label{evolution}

The very high luminosity of GR~290 places the star in the upper H-–R 
Diagram near the most luminous early--type stars. Presently GR~290 has 
many spectral characteristics in common with the late WN stars 
displaying hydrogen lines, 
except for the large spectrophotometric variability. 
In the evolutionary sequence of very massive stars the position of 
stars with a spectrum similar to late WN--type and with 
hydrogen in their spectra, the so--called WNH stars (Smith \& Conti 2008), 
is still under debate. They may be in an early evolutionary phase of 
core He burning with an H envelope not yet completely dissipated. 
In this scenario, the WNH phase occurs immediately after, or instead of, the 
LBV phase. Another school of thought puts at least the most luminous WNH 
stars (log $L/L_{\odot}$ above 5.8--6.0) before the LBV phase, so in a core H burning 
stage, evolving directly from the main sequence to the Wolf-–Rayet stage 
perhaps even without an intermediate LBV phase. 
The masses of such WNH stars seem to be statistically higher than 
genuine hydrogen--free Wolf--Rayet stars (Smith \& Conti 2008), 
and this may indicate that they are much less evolved objects. 
The following LBV phase is then consequence of their high luminosity 
with mass loss rates determined by the Eddington luminosity limit being exceeded.  
GR~290 has a luminosity around 10$^6$~$L_{\odot}$, therefore of the same 
order as expected for the very luminous WNH stars, 
except that displays large long term photometric variations typical of LBVs, 
while WNH stars do not. 
An evolutionary development has been suggested in the literature with very 
massive stars with initial mass larger than $\sim$40 $M_{\odot}$  
displaying LBV activity very early in their evolution, 
perhaps even while in core H burning. 
During this stage the stars would still be very luminous and display a 
variable WN9--11 spectrum while undergoing the LBV transient $hot$ phases,
as indeed observed for GR~290, before progressing on to WN8 stars 
(see Smith et al. 1994, Crowther et al. 1995, Crowther \& Smith 1997). 
GR~290 is a typical LBV for its large spectrophotometric variations and, 
presently, it is in a state, rather extreme for an LBV, 
contiguous to the location of the WN8 stars. 
Additionally, from the spectrophotometric behaviour observed in recent years, 
we cannot exclude that the star is going through a transition phase, 
perhaps developing after some time the expected WN8 spectrum.


\section{Conclusions} \label{conclusion}

We have used recent spectrophotometric data together with existing literature data 
to study the long--term behaviour of the LBV star GR~290.   
The star is peculiar because of the high excitation temperature at minimum, 
higher than ever observed in a confirmed LBV. 
We find that one cannot easily account for the observed spectral and luminosity 
changes without assuming that the bolometric luminosity has significantly changed 
during the present minimum phase, 
in contrast with what is generally believed in the case of LBVs. 
There are few other LBV objects with such behaviour, 
which, as discussed above, can be attributed to 
conversion of radiative power into mechanical power. 
We advance the alternative hypothesis that the luminosity decrease 
may only be apparent and be due to an increase in circumstellar extinction. 
We also explore the possibility that, because of its very high luminosity 
and as indicated by its extreme spectrum, the star 
now is not too far to end this phase and to enter a late WN type star. 
This is of course only a hypothesis that in particular awaits further tests 
on surface chemical abundances, 
in order to determine the star's actual evolutionary stage. 

\acknowledgments

Thanks are due to the TAC of the Isaac Newton Group of 
Telescopes for allocating service time at the WHT telescope, which was 
fundamental for our work.  
We are grateful to Phil Massey for having put at our disposal
the September 2006 spectrum of GR~290 for analysis,  
and to Alessandro Chieffi for useful discussions. 
We would like to thank the anonymous referee for constructive comments 
and for having drawn our attention to the preprint of the article on GR~290  
by Maryeva and Abolmasov. 
We also wish to ackowledge the Roma Amateur Astronomy Association (ARA) 
and Franco Montagni for undergoing on our behalf a monitoring program to
alert possible luminosity variations of GR~290 and other peculiar objects. 



{\it Facilities:} \facility{WHT (ISIS)}, \facility{Bologna 
Cassini Telescope (BFOSC)}, \facility{Cima Ekar (AFOSC)}.

\begin{figure}
\epsscale{1.00}
\plotone{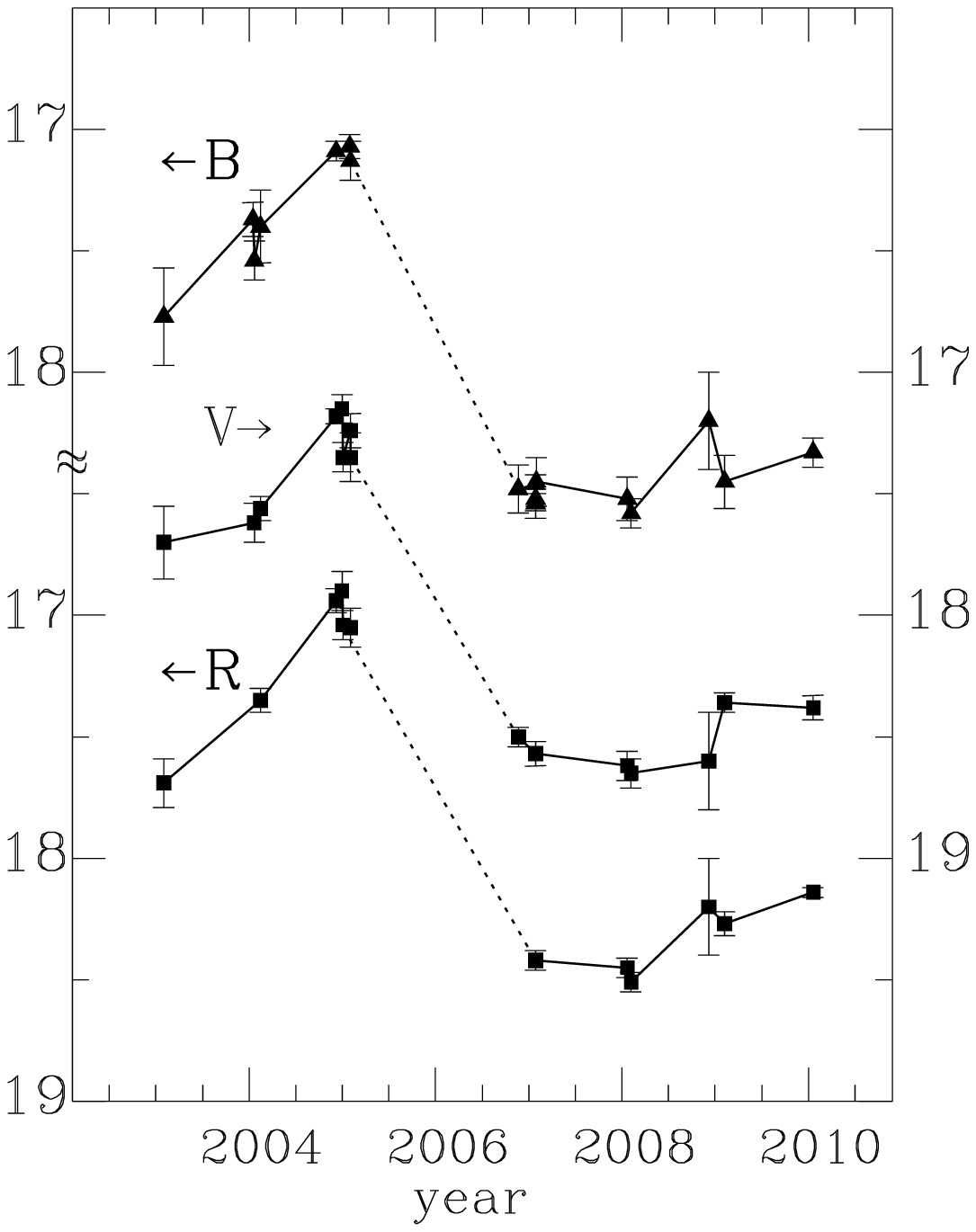}
\caption{The recent light curve of GR~290 from February 2003 
to January 2010 in the $B$, $V$ and $R$ bands. 
For clarity, the different curves have been vertically shifted.  
As indicated by arrows next to $B$~$V$~$R$ labels in the plot, 
one reads the $B$ magnitudes in the upper part 
of the left scale and the $R$ magnitudes in the lower part, while  
the right scale refers to the $V$ magnitudes.\label{fig1}}
\end{figure}


\begin{figure*}
\epsscale{2.00}
\plotone{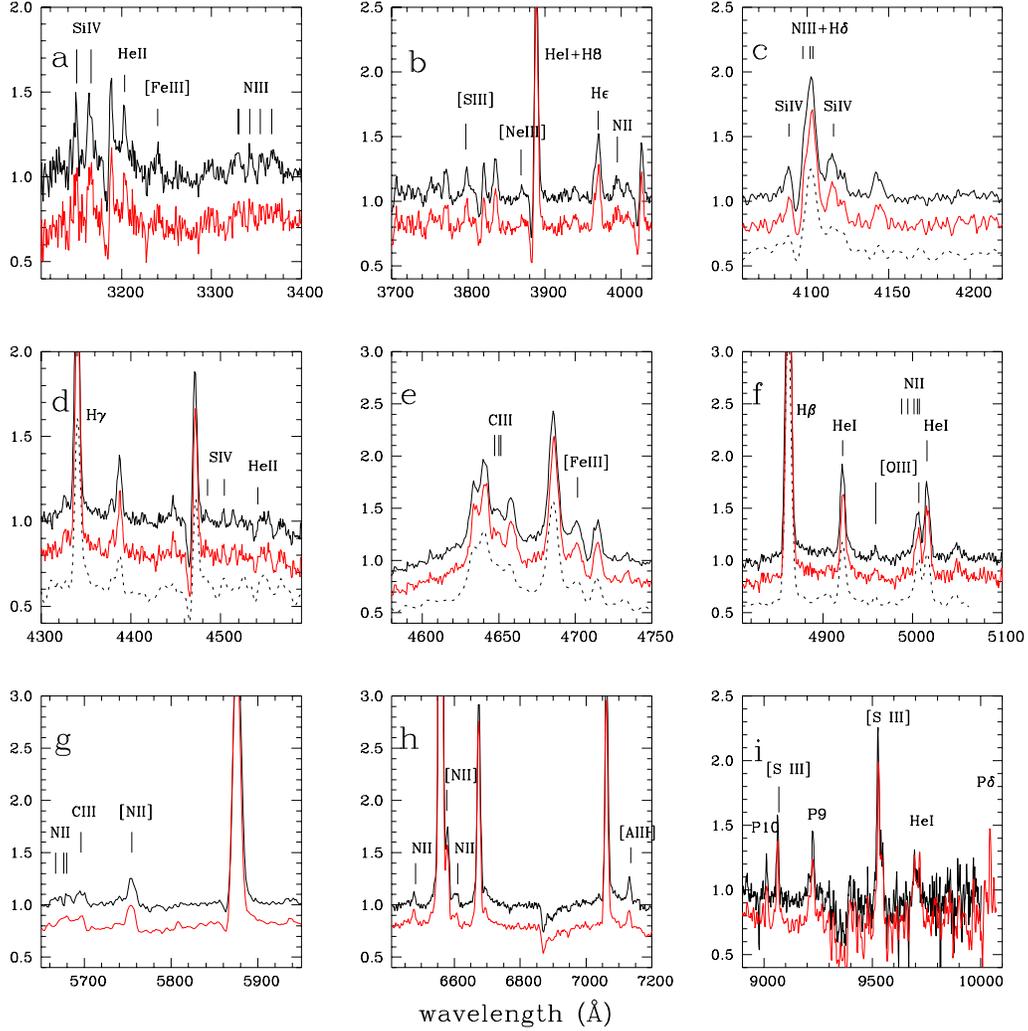}
\caption{
2008 spectrum of GR~290 with some line identification.
Two expsures are shown for each spectral region, 
with a vertical offset of $-$0.2 for the less exposed one (in red), 
to aid with assessing the presence of the various spectral features. 
In panels $c$-$f$ the WIYN spectrum of September 2006 is also shown (dots) 
with a vertical offset of $-$0.4. 
Some lines of interest are identified in the figure. 
The vertical scale represents fluxes normalized to the continuum.
The wavelength scales of the spectra have been shifted to fit the 
laboratory values.\label{fig2}}
\end{figure*}


\begin{figure}
\epsscale{1.00}
\plotone{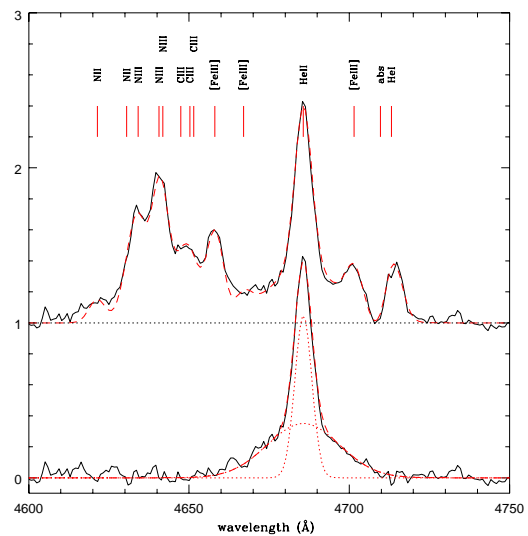}
\caption{
$Upper~panel$: The 4620--4713 \AA\ spectral range, as observed in 
December 2008, fitted (dashed line) by a combination of selected emission lines 
(marked by vertical bars and identified in the picture). 
The wavelength scale of the spectrum has been shifted 
to fit the laboratory wavelength. 
Both broad and narrow components have been included for the He \II\ 4686 \AA\ line 
(see lower panel). 
For the He \I\ 4713 line a violet--shifted absorption component 
has been included in the fit. 
$Lower~panel$: 
Spectral residual distribution around the He \II\ 4686 \AA\ line, after subtraction 
of the contribution of all the selected lines except He \II, 
fitted (dashed line) by narrow and broad Gaussian profiles (dotted lines).\label{fig3} 
}
\end{figure}


\begin{figure} 
\epsscale{1.20}
\plotone{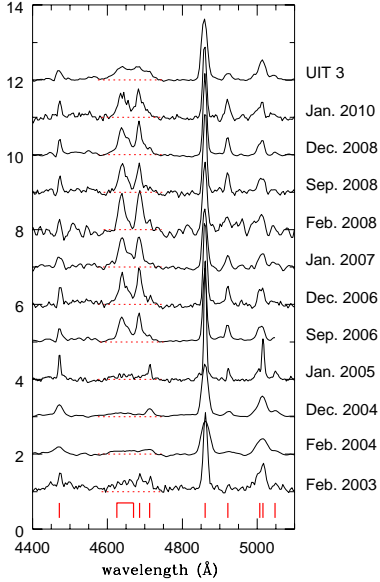}
\caption{Spectral variation of GR~290 during 2003--2010. 
Ordinates are fluxes normalized to the continuum, 
with vertical offsets. 
As for comparison, we show on the top the Cima Ekar December 2004 
spectrum of the Of/WN9 star UIT~3 in M~33 (see Viotti et al. 2006). 
The WIYN and WHT spectra of September 2006 and December 2008 
have been degraded to the resolution of the Loiano spectra. 
The Cima Ekar 2004 spectra of GR~290 and UIT~3 
have a slightly lower spectral resolution than those taken with the 
Loiano telescope, so that the 4630--4660 \AA, He \II\ 4686 \AA\ 
and He \I\ 4713 \AA\ emissions are less resolved.  
The vertical bars at the bottom mark the following emission lines: 
He \I\ 4471 \AA, the 4630-4670 \AA\ blend, 
He \II\ 4686 \AA, He \I\ 4713 \AA, H$\beta$, He \I\ 4922 \AA, 
[O \III]+N\II\ 5007 \AA, He \I\ 5016 \AA, and He \I\ 5048 \AA.\label{fig4}
} 
\end{figure}




\begin{figure}
\epsscale{1.00}
\plotone{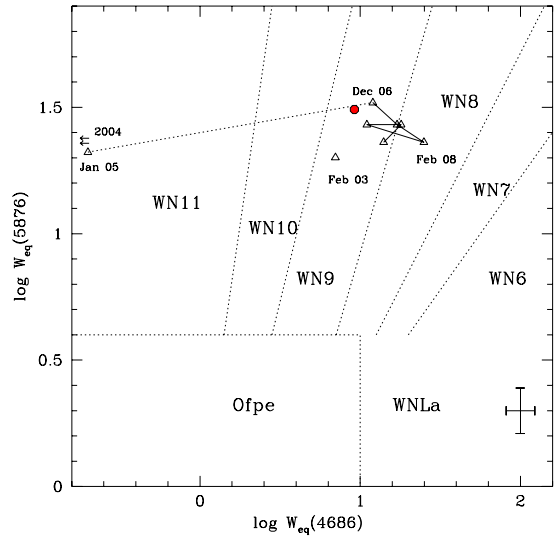}
 \caption{Log equivalent widths of He \I\ 5876 \AA\ 
versus He \II\ 4686 \AA\ for GR~290 during 2003--2010. 
The arrows mark the 2004 observations when the He \II\ line 
was not measurable. 
The dotted lines mark the approximate boundaries of different spectral classes 
according to Crowther \& Smith (1997) for galactic and LMC stars. 
The error bar of the measurements is shown on the lower right. 
The filled circle indicates the position 
of the Of/WN9 star UIT3 in M~33.\label{fig5}}
\end{figure}










\begin{figure}
\epsscale{1.20}
\plotone{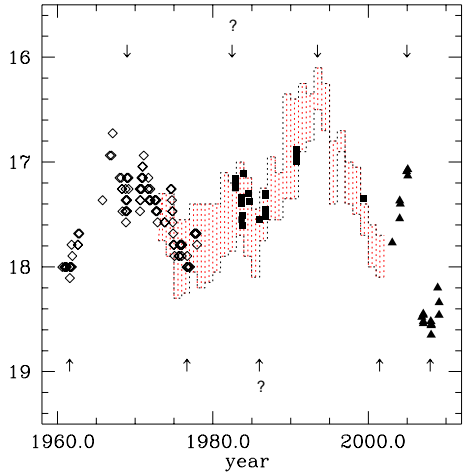}
 \caption{The historical light curve of GR~290 in the $B$--band, 
based on the photographic 1962--1978 survey of Romano (1978, diamonds), 
on the photographic 1982--1990 monitoring and $B$--survey (June 1999)  
of Kurtev et al. (2001, filled squares), 
and on our 2003--2010 $B$--monitoring (filled triangles). 
The hatched area contains the photographic  
Sternberg Astronomical Institute and Baldone observations 
obtained from 1972 to 2000, and reported by Sharov (1990),  
Sholukova et al. (2002), and Maryeva \& Abolmasov (2010). 
The original photographic observations 
have been converted into Johnson's $B$ magnitudes using the 
relation: $B$~=~1.064~m$_{ph}$~$-$0.831 (see Kurtev et al. 
2001, and Sholukhova et al. 2002). 
The proposed times of the light minima and maxima 
are indicated by arrows.\label{fig6}}
\end{figure}



\begin{figure}
\epsscale{1.00}
\plotone{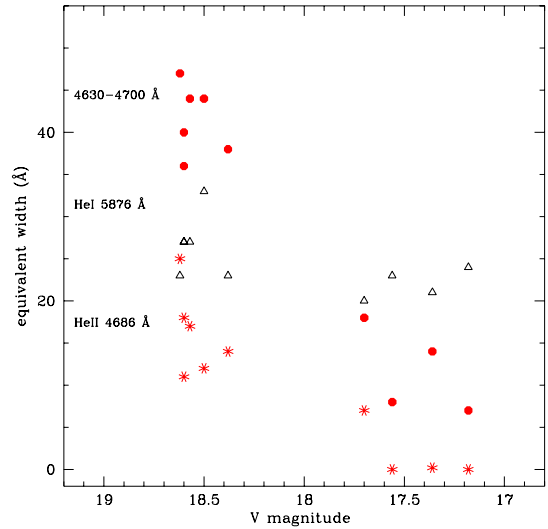}
\caption{ 
GR~290: the spectrum--luminosity anticorrelation during 2003--2010. 
The equivalent widths of He \II\ 4686 \AA\ (stars), 
He \I\ 5876 \AA\ (open triangles), and of the 
4630--4700 \AA\ blend (stars), are plotted against 
the visual magnitude of GR~290.\label{fig7}}
\end{figure}



\begin{table*}
\caption{New photometry of GR~290.\label{table1}}
\begin{tabular}{llllll}
\tableline\tableline
date           &~~~ $B$    &~~~ $V$    &~~~$R$     &~~~$I$ & telescope  \\
\tableline
2007 Jan. 28   &18.48(.05) &18.57(.05) &18.42(.04) &          & Loiano \\
2008 Feb. 07   &18.52(.07) &18.62(.04) &18.41(.09) &          & Loiano \\
2008 Sep. 08   &           &           &18.2(.1)   &          & Loiano \\ 
2008 Dec. 06   &18.2(.2)   &18.6(.2)   &18.2(.2)   &          & Cima Ekar  \\
2009 Feb. 09   &18.45(.11) &18.36(.04) &18.27(.05) &18.40(.09)& Loiano \\
2009 Oct. 26   &18.33(.06) &18.33(.04) &18.16(.03) &          & Loiano \\
2010 Jan. 21   &18.33(.06) &18.38(.05) &18.14(.03) &          & Loiano \\  
\tableline
\end{tabular}
\end{table*}






\begin{deluxetable}{lrll}
\tablewidth{0pt}
\tablecaption{The December 2008 WHT spectrum of GR~290.\label{Table2}}
\tablehead{
\colhead{$\lambda$\tablenotemark{a}}    & \colhead{\weq\tablenotemark{b}}   &
\colhead{ident.}          & \colhead{remarks\tablenotemark{c}} }
\startdata
3147    &-1.9:  & Si \IV\ 49.56  & \\	
3161    &-2.6:  & Si \IV\ 65.72  & \\ 
3181.0a & ~1.4  & He \I\  87.74  & PC \\
3185.1  &-1.9   & He \I\  87.74  & \\
3200.8  &-0.7   & He \II\ 03.04  & \\ 
3238.8  &-1.1   & [Fe \III] 39.74& \\
3326.7  &-0.8   & N \III\ 29.49  & bl N \III\ 30.11 \\ 
3335.8a &~0.2   & N \III\ 42.78  & PC \\ 
3341.9  &-0.6   & N \III\ 42.78  &  \\
3348.5a & ~0.3  & N \III\ 54.33  & PC\\
3353.7  &-0.7   & N \III\ 54.33  & \\
3359.1a & ~0.3  & N \III\ 67.25  & PC \\
3366.8  &-1.8   & N \III\ 67.25  & bl 3374\\
3582.2a & ~0.49 & He \I\ 87.25   & PC \\
3585.3  &-0.41  & He \I\ 87.25   & \\
3627.8a & ~0.6  & He \I\ 34.23   & PC \\
3632.6  & -1.1  & He \I\ 34.23   & \\
3697.7a & ~0.3  & He \I\ 05.00   & PC \\
3703.9  &-0.6   & He \I\ 05.00   & \\
3748.7  &-1.1   & H12 50.15      & ct N \III\ 52.63\\
3758.6  &-0.7   & N \III\ 62.60  & weak \\ 
3768.6  &-1.4   & H11 70.63      & ct N \III\ 3771 \\
3795.3  &-1.0   & [S \III] 96.7  & \\ 
3801.9  & -0.6  & Si \III\ 06.56 & \\ 
3812.6a & ~0.85 & He \I\ 19.61   & PC \\
3818.1  &-1.1   & He \I\ 19.61   & \\
3833.0  &-2.1   & H9 35.39       & \\ 
3866.9  &-0.5   & [Ne \III] 68.74& \\ 
3881.4a & ~2.4  & He \I\ 88.65   & PC \\ 
3886.4  &-10.9  & He \I\ 88.65   & ct H8 3889\\ 
3961.7a & ~0.7  &                & PC H$\epsilon$, He \I\ 3964\\ 
3967.7  &-3.75  & H$\epsilon$ 70.07& bl He \I\ 64.73\\ 
3992.5  &-1.18  & N \II\ 95.00   & \\ 
4005.7  &-0.45  & He \I\ 09.27   & \\ 
4019.1a & ~1.1  & He \I\ 26.19   & PC \\ 
4024.3  &-2.3   & He \I\ 26.19   & \\ 
4086.4  &-1.7   & Si \IV\ 88.86  & \\ 
4090.8a &-1.0   &                &PC H$\delta$, N \III\ 4097 \\
4095.0  &-2.6:  & N \III\ 97.31  & bl 4101 \\
4099.8  &-8.5   & H$\delta$ 01.74& bl N \III\ 03.37 \\ 
4112.3  &-2.1   & Si \IV\ 16.10  & \\ 
4119.1  &-1.2   & He \I\ 20.82   & bl 4116 \\ 
4140.8  &-1.2   & He \I\ 43.76   & \\ 
4324.0  &-0.6    & N \III\  27.69 ?& bw H$\gamma$ ? \\
4337.5  &-6.8    & H$\gamma$ 40.47 & \\ 
4375.6  &-0.8    &                 & bw 4387 ?  \\ 
4385.4  &-2.2    & He \I\ 87.93    & \\ 
4444.8  &-0.7    &                 & n.i. \\ 
4463.5a & ~2.1   & He \I\ 71.48    & PC \\ 
4469.0  &-5.5    & He \I\ 71.48    &  \\ 
4481.8  &-0.2    & S \IV\ 85.66    & W\&R \\
4501.0  &-0.5    & S \IV\ 04.09    & W\&R \\ 
4511.3  &-0.6    & N \III\ 14.85   & \\ 
4536.3a &~0.5    & He \II\ 41.59   & \\ 
4546.4: &-2.2:   & Si \III\ 52.65  & \\ 
4565.5: &-1.1:   & Si \III\ 67.87  & \\ 
4630.5  &-4.0    & N \III\ 34.14   & \\ 
4637.3  &-6.2    & N \III\ 40.64   & bl N\III\ 41.90 \\ 
4645.6: &-4:     & C \III\ 47.40   & bl C\III\ 50.16, 51.35\\
4654.4  &-3.8    & [Fe \III] 58.05 & \\
4682.0  &-17     & He \II\ 85.68   & narrow+broad\\ 
4697.9  &-1.7    & [Fe \III] 01.53 & \\
4705.1a &~1.3    & He \I\ 13.15    & PC \\ 
4709.5  &-3.0    & He \I\ 13.15    & \\ 
4858.2  &-18     & H$\beta$ 61.33  & \\ 
4918.7  &-2.2    & He \I\ 21.93    & \\ 
4954.6  &-0.7    & [O \III] 58.91  & \\ 
5002.9  &-3.4    & [O \III] 06.84  & ct N \II\ m.19,24\\
5012.5  &-4.5    & He \I\ 15.68    & \\ 
5044.3  &-1.4    & He \I\ 47.74    & \\ 
5264.9  &-2.8    & [Fe \III] 70.42 & \\ 
5672.2  &-1.5    & N \II\ 79.56    & bl N \II\ 76.02 \\ 
5690.0  &-1.6    & C \III\ 95.92   & \\ 
5748.3  &-3.1    & [N \II] 54.8    & \\ 
5865.7a &~1.9    & He \I\ 75.62    & PC \\ 
5871.9  &-29     & He \I\ 75.62    & \\ 
6307.2  &-1.6    & [S \III] 12.06  & He \I\ 10.83? \\ 
6476.3  &-1.0    & N \II\ 82.07    & \\ 
6540.5  &-2.9    & [N \II] 48.1    & \\ 
6557.9  &-121    & H$\alpha$ 62.82 & \\ 
6578.1  &-8.5    & [N \II] 83.6    & \\ 
6604.5  &-2.4    &  N \II\ 10.58   & bl ?  \\ 
6671.8  &-25     & He \I\ 78.12    & \\ 
6693.6& -1.2     &  n.i.           &n.i.  bl 6678\\ 
7050.0a &~1.3    & He \I\ 65.19    & PC \\ 
7059.8  &-25     & He \I\ 65.19    & \\ 
7129.3  &-2.4    & [A \III] 35.8   & \\ 
7275    &-7      & He \I\ 81.35    & \\ 
7319    &-2.5    & [O \II] 19.91 ? & \\ 
7742    & -0.6   & [A \III] 51.5?  & \\ 
7882    & -3     &                 & n.i., double \\ 
8660    &        & P13 65.02       & \\ 
8747    &        & P12 50.48       & \\ 
8858    &        & P11 62.79       & \\ 
9010    &        & P10 14.91       & \\ 
9060    &        & [S \III] 69.4   & strong\\		
9222    &        & P9 29.02        & strong\\
9528    &        & [S \III] 32.1   & strong \\
9542    &        & P8 45.97        & bl 9532 \\ 
9698    &        & He \I\ 02.66    &  \\
10022   &        & P$\delta$ 49.38 &  \\
\enddata 

\tablenotetext{a}{Observed wavelengths. $a$: absorption line.}
\tablenotetext{b}{Equivalent widths in \AA, negative for emission lines.}
\tablenotetext{c}{Remarks:  $n.i.$: not identified. 
$bw$: possibly blue wing of the nearby line cut by its P~Cygni absorption component. 
$PC$: P~Cygni absorption component. 
$ct$: contributing line. 
$bl$: blended with nearby line. $m$: multiplet number. 
$W\&R$: lines identified by Werner \& Rauch (2001). }


\end{deluxetable}

\clearpage


\begin{table*}
\caption{The variable spectrum of GR 290\tablenotemark{a}\label{table3}}
\begin{tabular}{lccrcrrrrl}
\tableline\tableline
date/target& $V$&sp. type\tablenotemark{b} & sp. type\tablenotemark{c}  
& H$\alpha$\tablenotemark{d} & 4630-4700\tablenotemark{e}& 4686& 5876& remarks\\
\tableline
2003 Feb. 02        & 17.70 & WN9   &   & 105 & 18  & 7     & 20  & Loiano\\
2004 Feb. 14        & 17.56 & WN11  &   & 100 & ~8  & n.m.  & 23  & Cima Ekar\\
2004 Dec. 07        & 17.18 & WN11  &   & 118 & ~7  & n.m.  & 24  & Cima Ekar\\
2005 Jan. 13        & 17.36 & WN11  &   & 122 & 14  & 0.2   & 21  & Loiano\\
2006 Dec. 14        & 18.50 & WN9   &WN9& 135 & 44  & 12    & 33  & Loiano\\
2007 Jan. 29        & 18.57 & WN8-9 &   & 126 & 44  & 17    & 27  & Loiano\\ 
2008 Feb. 07        & 18.62 & WN8   &   & 129 & 47  & 25    & 23 & Loiano\\ 
2008 Sep. 08        &(18.6)~& WN9   &   & 120 & 36  & 11    & 27  & Loiano\\ 
2008 Dec. 04        & 18.6~ & WN8-9 &WN9& 130 & 40  & 18    & 27  & WHT   \\ 
2010 Jan. 21        & 18.38 & WN9   &   & 120 & 38  & 14    & 23  & Loiano \\
\tableline
\end{tabular}
\tablenotetext{a}{New and revised old equivalent widths 
of the emission lines in \AA.}
\tablenotetext{b}{Equivalent spectral types for GR 290 according to the diagram 
of Crowther \& Smith (1997).}
\tablenotetext{c}{Spectral types from the mid resolution WIYN and WHT spectra.} 
\tablenotetext{d}{Including the line wings and the [N \II] lines.}
\tablenotetext{e}{The blend includes [Fe \III] 4701 \AA. 
He \I\ 4713 \AA\ is excluded.}
\end{table*}




\end{document}